# Spatial characterization of the edge barrier in wide superconducting films

A. G. Sivakov, O. G. Turutanov [a)], A. E. Kolinko, and A. S. Pokhila

*B.I. Verkin Institute for Low Temperature Physics and Engineering, NAS of Ukraine, 47 Nauky Ave., Kharkiv 61103, Ukraine*

The current-induced destruction of superconductivity is discussed in wide superconducting thin film strips, whose width is greater than the magnetic field penetration depth, in weak magnetic fields. Particular attention is paid to the role of the edge potential barrier (the Bean-Livingston barrier) in critical state formation and detection of the edge responsible for this critical state with different mutual orientations of external perpendicular magnetic field and transport current. Critical and resistive states of the thin film strip were visualized using the space-resolving low-temperature laser scanning microscopy (LTLSM) method, which enables detection of critical current-determining areas on the thin film edges. Based on these observations, a simple technique was developed for investigation of the critical state separately at each film edge, and for the estimation of residual magnetic fields in cryostats. The proposed method only requires recording of the current-voltage characteristics of the thin film in a weak magnetic field, thus circumventing the need for complex LTLSM techniques. Information thus obtained is particularly important for interpretation of studies of superconducting thin film single-photon light emission detectors.

[a)] E-mail: turutanov@ilt.kharkov.ua

**1. Introduction**

Over the last few decades, superconductivity suppression, critical currents, and resistive state in 1D and 2D superconductors have attracted increased interest from theorists and experimenters, whose work has enabled significant progress in these areas of research. However, currently, research efforts are focused on resolving the problem of the destruction of superconductivity in the intermediate case of relatively narrow thin films, in connection with the development of new superconducting single-photon detectors (SSPD)[1] of visual and infrared light based on ultrathin (1–10 nm) films of NbN, MoSi, MoRe, and other similar superconducting materials with small (one nanometer) coherence length. Nanowire width in SSPD is such that they could be attributed not to quasi-1D, but to quasi-2D superconductors. For these applications, a clear understanding of the critical current local mechanisms, effects of the width and geometry of film edges including edge non-homogeneity, and weak (residual) magnetic fields. It is known that, in quasi-1D wires, destruction of superconductivity occurs due to thermally activated[2–4] and transport-current-generated[5–8] phase-slip centers (PSCs) of the order parameter, which are responsible for further development of the resistive state.

In quasi-2D superconducting thin film systems, the development of resistivity is attributed to thermally activated vortex-antivortex pairs[9–11] in the superconducting transition region and the occurrence of Abrikosov vortices[12] of an external magnetic field or self current-induced field. In the case of weak volume pinning of vortices, low critical current could be expected in the film, however, vortices entrance is prevented by potential energy barrier (the Bean-Livingston barrier[13]) formed at the edge of the film, which determines the critical current magnitude.

It is well known that, in "wide" films with non-uniform supercurrent distribution over the cross section, any edge defects, such as notches, sharp cross section narrowing and even broadening, lead to a decrease in the critical current. The strong effect exerted by edge geometry on critical current resulted in the introduction of the "geometric barrier" concept[14,15]. Glover and Coffey[16] demonstrated that, taking into account non-uniform transverse current density distribution in the film strip, current-induced destruction superconductivity occurs when the Ginzburg-Landau depairing current density is reached at the edge[17]. Within the framework of the Ginzburg-Landau theory, a unified approach was applied to superconductors of any width. This enables the determination of the superconductor critical state by searching for the most probable trajectory of a system transition between metastable states within the configuration space through the saddle points of the Ginzburg-Landau functional. The probability of these transitions is determined by the energy barrier height. The behavior of the superconductor changes with its width $w$=4.4$\xi$, which separates the 1D domain with phase-slip type solution and a 2D domain with vortex type solution. Nevertheless, it is the edge barrier that determines the critical state onset, regardless of the width and dimension of the superconductor. Moreover, it has been shown[18] that the size and shape of the edge defect do not affect the value of the critical current, even in the case of a quasi-1D channel with uniform distributions of current and order parameter.

These previous works share a common edge energy barrier concept which determines the boundary of the superconducting state stability. Previously[19], we visualized the edge barrier in wide superconducting tin films using low-temperature laser scanning microscopy (LTLSM) technique[20]. In the present study, we decided to revisit the important role of the edge barrier issue, as previous works have largely ignored the fact that the critical current is determined by one of the edges of the superconductor; but which one works is generally not known, and this may also be determined by the weak residual magnetic field frozen in a cryostat. In particular, this makes it difficult to understand the mechanism by which individual photons are detected in superconducting thin film detectors of visual and infrared light. In this study, using the space-resolving LTLSM technique, we demonstrate spatial imaging of current-induced destruction of superconductivity in tin films, simulating the processes in nanoconductors used in SNSPD. (It is not possible to visualize the resistive state in real nanowires using this method because of the fundamental limitations on the spatial resolution by the light wavelength). On the basis of the obtained results, we propose a simple technique for estimating the state of both edges, the asymmetry of the edge barrier, and residual magnetic field trapped in the cryostat, that includes recording of current-voltage characteristics (I–V curves, IVCs) in weak magnetic field. This technique is facile and does not require a complex LTLSM setup.

## 2. Experimental technique

The samples in our experiment were thin (30–50 nm) tin films formed by various methods (scribing, laser cutting, electron lithography) in the form of 20–100-µm-wide strips with potential output leads and expanding current input leads. Large coherence length and magnetic field penetration depth in tin films, compared with the corresponding parameters for superconducting compounds such as NbN, MoSi, and MoRe, made it possible to simulate situation with current-induced destruction of superconductivity in the latter materials at much larger strip widths than in the original nanowires of these materials. This enabled the use of LTLSM with about one-micron spatial resolution for visualization of critical state formation and transition of films to the resistive state. The basis of the LTLSM method and its application to the investigation of local critical currents and imaging of the resistive state in wide films are described in detail in our review[20]. It should be noted that the sample response in this method is the change in voltage $\delta V$ at the ends of the sample in the resistive state or during film transition from superconducting state to resistive state which caused by suppression of superconductivity due to locally irradiation with the light probe at a given transport current. As a rule, beam intensity is modulated at a frequency of 1–100 kHz, and the alternating response voltage $\delta V$ is amplified and detected by the lock-in amplifier. Response amplitude is imaged by representing its spatial distribution in the film plane in the form of a 3D profile, halftone map, or transverse coordinate dependence curves in individual sections. Note that, for a wide film, response magnitude in an individual cross section as a function of the longitudinal coordinate (transport current flow direction), is inversely proportional to the local superconducting current value in this section. In addition, response amplitude across the sample (i.e., the response distribution over the cross section) is proportional to the local critical current density[21].

Recording of the current-voltage characteristics $V(I)$ and critical currents vs. magnetic field $I_c(H)$ was carried out in a PC-controlled automated setup. To determine critical current dependencies on external parameters, a set of IVCs was recorded and the critical current value was determined using software via the detection of a small specified voltage during automatic recording of the current-voltage characteristics, or during their subsequent viewing.

## 3. Mechanisms of resistivity in superconducting thin films

Consider a scenario in which the current-induced destruction of superconductivity occurs in wide film strips, with width $w$ exceeding both the temperature-dependent coherence length $\xi(T)$ and magnetic field penetration depth $\lambda(T)$ (more precisely, the value of $\lambda_\perp(T) = 2\lambda^2(T)/d$, where $d$ is the film thickness), in zero external magnetic field. Such large film width assumes a vortex resistivity mechanism. In particular, the role of the edge barrier in the formation of the critical, and then resistive state should be noted here. One of the first studies explaining the main features of current-induced destruction of superconductivity and development of the resistive state up to the current jump on the current-voltage characteristic was the Aslamazov and Lempitsky theory[22]. This theory was based on the analysis of the Meissner state instability conditions influenced by infinitesimal perturbations of order parameter and vector potential, leading to edge barrier suppression and penetration of vortices into the film, within the framework of the Ginzburg-Landau theory approximation. In this model, complete suppression of the Bean-Livingston barrier occurs when current density at the edge reaches the Ginzburg-Landau depairing current density, with spatial periodicity of critical instability along the film edge. As soon as critical current is reached and the barrier is suppressed, vortex line ("rank") of different signs, enter the film from opposite edges and move towards each other, annihilating at the central line. Vortices contribute to the current density, which has maximum at the central line of the film strip. The maximum value, according to the numerical calculations[23] for moderately wide films, increases linearly with increasing transport current until it reaches the Ginzburg-Landau depairing current density. With this current, instability of the stationary vortex flow arises, which is expressed at break point (voltage jump) on the IVC. An alternative explanation of the break point is given by a previously described theory[24], which takes into account nonlinearity of the vortex dynamics (viscosity decreases when vortices move with increasing velocity). Further evolution of the resistive state is not considered by both theories, implicitly assuming that transition follows to the normal state.

An experimental study[25] showed that, for sufficiently good heat sink at high currents corresponding to the break point in the theories[22,24], a transition occurs to another resistive state, known as phase slip lines (PSLs), which is a generalization of the concept of phase slip centers (PSCs) in narrow superconducting channels. In this paper, the resistive state with PSL is not discussed. We will focus on analyzing critical state formation induced by the transport current and magnetic field, and its spatial localization.

## 4. Results and discussion

Figs. 1(a) and 1(b) present 3D maps of the resistive LTLSM response of a current-carrying superconducting tin film for two specific values of the transport current.

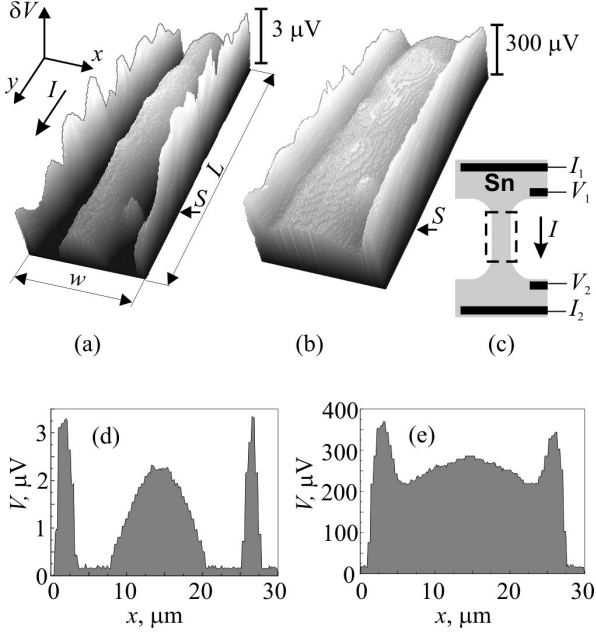

Fig. 1. Imaging of resistive state of the wide superconducting tin film in zero magnetic field using LTLSM. 3D map of voltage response $\delta V$ ("mountain landscape") of rectangular working part of the sample at the transport current (a) $I \cong I_c$, $I > I_c$ and (b) $I \gg I_c$. The width of the scanning area $w = 30\,\mu m$, the length $L = 100\,\mu m$. The arrows indicate transport current ($I$) direction and selected sections ($S$) for 1D distributions of the response across the film (d), (e). Configuration of the film sample is demonstrated (c) with current and potential leads; the dashed line limits the scanned area.

The film with a thickness of 30 nm was deposited in vacuum on a crystalline quartz substrate. Figure 1(c) shows the sample configuration; size of the scanning area (the working part of the film with uniform cross-section) is also indicated. Expanding current input leads are not included in the scan area. Figs. 1(d) and 1(e) demonstrate corresponding responses vs. transverse coordinate in one of the sample cross sections, as indicated by the arrows and the letters S in Figs. 1(a) and 1(b), respectively.

When the transport current is slightly higher than the critical current ($I \approx I_c$, $I > I_c$), an almost periodic change is observed in the amplitude of the LTLSM response along the film (Fig. 1(a)). It is inversely proportional to the critical current density in the cross section and reflects the periodic variation of the edge barrier. Response maxima correspond to the places with suppressed barrier, i.e., the points where vortices enter the film. At the same time, supercurrent distribution across the film shows three maxima, two at the edges and one at the center (Fig. 1(d)). Note that the response is proportional to the local supercurrent density in transverse direction.

As the transport current increases, local supercurrent density distribution over the cross section becomes more and more uniform (Figs. 1(b) and 1(d)), approaching the depairing current density (note that voltage response scales for two response maps shown (a) and (b) differ by a factor of 100 due to differential resistance increase in the IVC prior to the jump to the PSL state).

Such an image is an experimental confirmation of the theoretical conclusions[22] and calculations[23], and completely fits the scenario described above. Due to the good quality of the edges of the electron-lithographed film (absence of sharp irregularities) and small value of the residual magnetic field in the cryostat, the barriers at both edges are almost identical.

In the absence of an external magnetic field ($H = 0$) the above-described picture is axially symmetric for identical edge barriers, and the magnetic fields of the current at the edges, $H_{IL}$ and $H_{IR}$ are equal in magnitude and opposite in sign, $H_{IL} = -H_{IR}$. Application of the external magnetic field $H$ breaks the symmetry of the barriers. Figs. 2(a)–2(c) demonstrates the results of the potential barrier imaging in a wide current-carrying film in zero and non-zero external perpendicular magnetic fields of opposite signs.

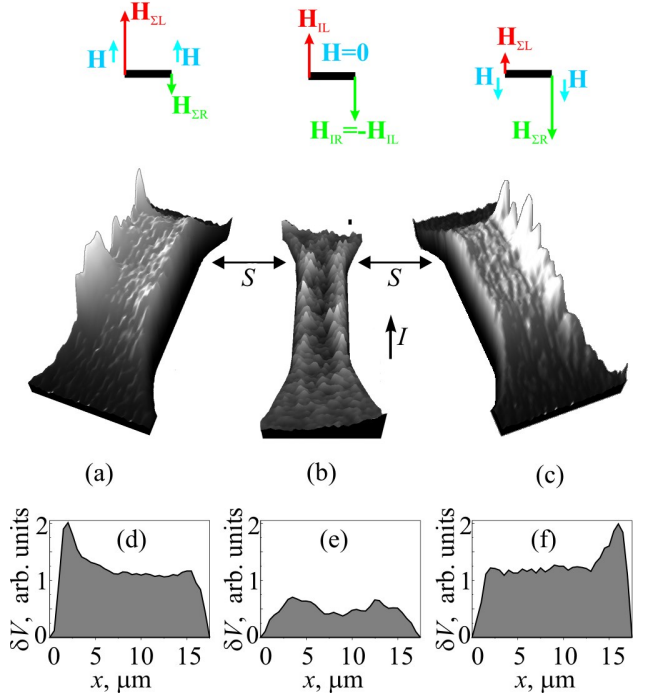

Fig. 2. Effect of external magnetic field on superconducting current density distribution and height of the edge barrier. 3D LTLSM map of the voltage response $\delta V$ of the superconducting tin film in the pre-critical state ($I \approx I_c(H=0)$, $I < I_c(H=0)$) in the external perpendicular magnetic field $H$ with induction, mT: -0.06 (a), 0 (b), and +0.06 (c). Arrows indicate the transport current ($I$) direction and selected sections ($S$) for 1D distributions of the response across the film (d), (e). On the inset above, the scheme of addition of external field $H$ and the current field $H_I$ on the left ($L$) and right ($R$) edges of the sample is shown.

Figures 2(d) and 2(e) demonstrates the corresponding response vs. transverse coordinate in one of the sections of the sample marked with the arrows and the letters S in Figs. 2(a)–2(c).

With $H = 0$ and pre-critical value of the transport current $I$ ($I \approx I_c$, $I < I_c$), transition to the resistive state occurs locally only under the action of the laser probe, and the Meissner pre-critical state is observed — local current density is almost symmetrically increased at both edges of the film (Figs. 2(b) and 2(d); see also explanations, in the previous figure, of the relationship between response and current density). The current magnetic fields at film edges $H_{IL}$ and $H_{IR}$ are equal in magnitude and opposite in direction (shown above the response maps, Fig. 2). When the external field $H \neq 0$ is applied, it is added algebraically to the current field $H_{I(L|R)}$ on the left (L) and right (R) edges, $H_{\Sigma(L|R)} = H_{I(L|R)} + H$ (inset in Fig. 2, case $|H| < |H_I|$). This leads to a substantial redistribution of the current density over the cross section (Figs. 2(d) and 2(e)) and asymmetric change of the barrier height at the edges in opposite directions (Figs. 2(a) and 2(c)).

In such an asymmetrical situation, vortices begin to enter only from one edge and exit on another edge, while there is no additional current density maximum inside the film. Critical state arises at the edge where the barrier becomes lower and the supercurrent density is higher than at the opposite edge as a result of addition of the external field and the current field. This is discussed further below. Thus, with a fixed transport current, the application of an external magnetic field of either sign leads to changes in the critical state spatial localization as a function of the mutual direction of the external field and current field.

Characteristic magnetic field values that cause fundamental changes in the critical behavior of the film are comparable to the Earth's magnetic field. Even for relatively narrow films with a width of 1 μm, the transport current of 1 mA creates edge field induction of 0.06 mT (corresponding to the field strength of 0.6 Oe) equal to the Earth's field. (To estimate the current field at the edge of the film strip, the simple and practical off-system equation $H_{I\,edge}[\text{Oe}] = 0.2\pi I[\text{mA}]/w[\mu\text{m}]$ may be used which follows from a more strict expression written in system units[26].) Fields of such magnitude may be completely trapped by the structural components of the cryostat if magnetic shielding is missed or insufficient. The shielding efficiency may deteriorate, for example, due to the presence of optical windows in the cryostats. In addition, unsuitably located supply wires may create such fields.

This highlights the need for thorough magnetic shielding of the cryostat for resistive studies and practical applications of relatively wide (tens of $\lambda_\perp$) superconducting films.

Critical current exhibits spatial localization; therefore, for some problems in which it is important to know exactly where the critical state arises, it is not sufficient to measure critical current $I_c$ and its dependence on the perpendicular magnetic field $I_c(H)$ for one arbitrary current direction and one direction of applied magnetic field $H$. We show below the information that may be extracted by obtaining $I_c(H)$ dependencies for all combinations of field and current directions.

In subsequent sections, we shall distinguish between the situations with different field and current mutual directions, introducing notation for the $I_c(H)$ curves $I+H+$, $I+H-$, $I-H+$, $I-H-$. The same signs of the field and current mean that the directions of external field and current field coincide on the "weaker" edge.

Fig. 3 demonstrates experimental dependencies of the critical current of the wide tin film vs. perpendicular magnetic field $I_c(H)$ for two directions of the transport current $I$ and the field $H$. All curves coincide and have a maximum in zero magnetic field $H = 0$.

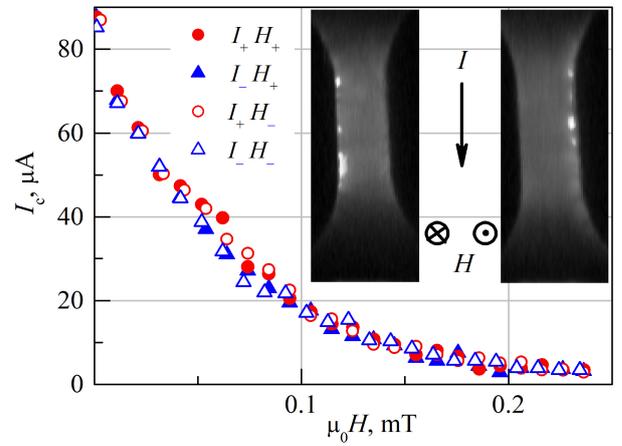

Fig. 3. Critical current vs. magnetic field for a film with symmetrical edge barriers. The density of the experimental points on the graph is reduced by a factor of 2 for better viewing. The legend denotes the mutual directions of the field and current (notations are explained in the text). Inset contains LTLSM maps of the film response for two opposite directions of the applied magnetic field $H$ with the same transport current $I$.

The LTLSM film response maps (see inset) show that, depending on the magnetic field sign for fixed current magnitude and direction, critical state arises on one or the other edge of the sample, although the measured critical current does not change. The same picture is observed if we fix field magnitude and direction, and change the sign of the transport current. The total symmetry with respect to the change in the directions of the field and current represents the identity of both edge barriers.

Such a case is an exception rather than a rule, as it is difficult to achieve complete identity of the barrier edges. Nevertheless, in the film studied, which was produced by electronic lithography, the quality of the edges was sufficiently good to ensure that the potential edge barriers were equal.

In the more common case, edge barriers are different due to the heterogeneity of the thickness, structure, edge geometry, and so on. The reasons for the differences between barriers are not important, and are not considered here. However, we note that critical current of the film for the vortex resistivity mechanism is always determined by the edge with a lower barrier. This case is illustrated in Fig. 4.

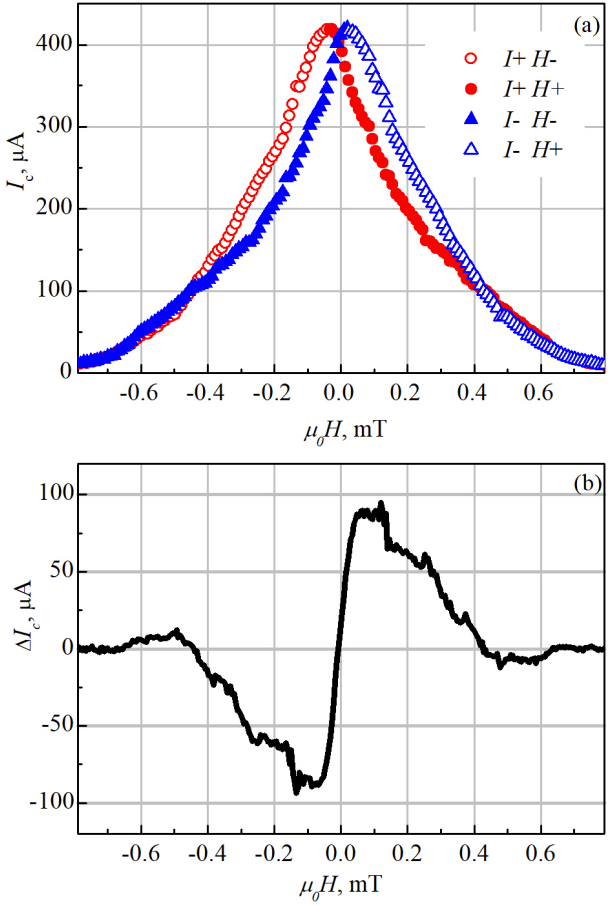

Fig. 4. Dependence (a) of the critical current $I_c$ and (b) of the difference of the critical currents from the magnetic field $H$ for a film with asymmetric barriers, with different directions of the transport current $I$. The density of the experimental points on the graph is reduced by a factor of 2 for better viewing.

For the positive direction of the transport current (experimental points $I_c$ are designated by circles), application of an external magnetic field $H$ in the direction coinciding with the field of current $H_I$ at the "weaker" edge (designated $I+H+$), suppresses the barrier at this edge, which leads to a decrease in critical current (the part of the curve with the filled circles in Fig. 4(a)). Simultaneously, on the opposite edge, the fields are subtracted, and the barrier grows until the external field compensates for the current field.

As the field $H$ increases in the other direction (case $I+H-$), it compensates for the current field $H_I$; the lower barrier increases for entry of vortices, and the critical current of the film increases. At the other edge, fields are added and the barrier to the entry of anti-vortices decreases. The critical current of the sample increases, reaching a maximum in the field at which the barriers become identical. With further increase in the field, the barriers skew to the other side, and other edge becomes "weak" and responsible for the critical current value, which begins to decrease (part of the curve indicated by empty circles).

The same reasoning is valid for the opposite direction of current (cases $I-H-$ and $I+H-$), herewith applied field direction should also be changed. As a result, for the negative direction of the transport current, we obtain the curve $I_c(H)$, which is mirror-symmetrical about the axis $H=0$ (curve with experimental points denoted by triangles). Figure 4(b) shows the difference in critical currents measured in opposite directions of the transport current, characterizing the barrier asymmetry. As shown, the curve $\Delta I_c(H)$ is completely symmetrical with respect to the center $I=0$, $H=0$. The fine structure of $\Delta I_c(H)$ may be a useful tool for the study of the physical mechanisms of critical current formation. The symmetry of the curves $I_c(H)$ and $\Delta I_c(H)$ with respect to $H=0$ indicates the absence of the residual (trapped) field $H_{res}$, and the symmetry $\Delta I_c(H)$ with respect to $I=0$ indicates independence of the barrier heights against magnetic field direction, but only against its absolute value.

For stronger fields in which the barrier disappears, critical current is determined not by the edge barrier, but by the volume pinning of the vortices, which does not depend on the direction of current and field; therefore, for strong applied fields, the curves merge (Fig. 4(a)), i.e., $\Delta I_c = 0$ (Fig.4(b)).

Thus, for an asymmetric barrier, we can determine the dependence $I_c(H)$ for each barrier separately.

Another factor that influences experimentally measured $I_c(H)$ dependencies, which was absent in our experiments, is residual field $H_{res}$ trapped in the cryostat. The residual field causes a shift of all curves $I_c(H)$ by the amount $-H_{res}$.

All possible experimental situations are shown schematically in Figs. 5(a)–5(d). In the upper row in Figs. 5(a)–5(d), $I_c(H)$ dependencies are shown in four quadrants for clarity, for two directions of the transport current and the applied perpendicular magnetic field. Curves referring to one specific (left or right) edge are conditionally denoted by the letters L and R.

The pictures in the bottom row correspond to the upper ones; however, the curves, as is customary in the experiment, are reduced to a positive half-plane (the critical current $I_c$ is a positive value). They better illustrate how dependence $I_c(H)$ varies in different cases when the measuring transport current direction changes (the direction to which the attention is often not paid in the experiment). Figure 5(a) illustrates the common case when edges are the same: the height of the edge barriers in zero field are equal, and there is no additional residual field. Under these symmetry conditions, the $I_c(H)$ dependence is independent on the directions of current and external applied field. An experimental example of this situation is shown in Fig. 3.

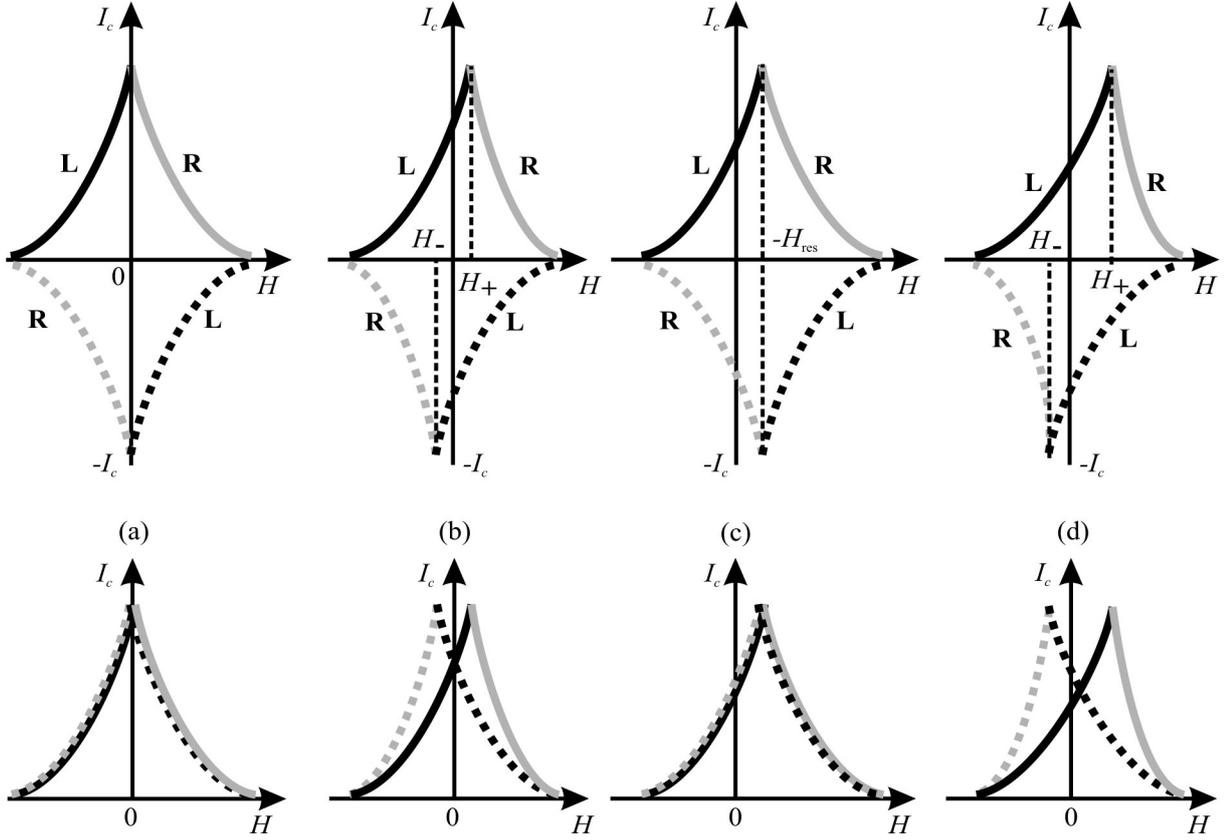

Fig. 5. Diagram showing the effects of edge barrier symmetry and trapped (residual) magnetic field $H_{res}$ on the critical current $I_c$ of wide superconducting film as a function of external magnetic field $H$ for different directions of the transport current $I$ and field $H$: (a) symmetric barriers, $H_{res}=0$, (b) asymmetric barriers, $H_{res}\neq 0$, (c) symmetric barriers, $H_{res}\neq 0$, (d) asymmetric barriers, $H_{res}\neq 0$. $H_-$ and $H_+$ are the values of applied field $H$, corresponding to the maxima of $I_c(H)$. Curves referring to one specific (left or right) edge are conditionally denoted by the letters L and R.

The case with initially asymmetric barriers in the absence of the trapped magnetic field (Fig. 5(b)) was considered in detail above for explanation of the experimental dependences in Fig. 4. The maxima $I_c(H)$ for different polarities of the transport current must be observed for different nonzero values of the field ($H_-$ and $H_+$), with $H_- = -H_+$, since $H_{res}=0$. The greater the asymmetry of the barriers, the greater is the absolute magnitude of the maxima shift over the field $|H_+|=|H_-|$. The set of curves $I_c(H)$ appeared to be symmetric with respect to the center of the coordinates.

If there is a residual magnetic field $H_{res}\neq 0$ in the cryostat, then, for the film, it is the same external field as the explicitly applied field $H$; therefore, all curves shift along the $H$ axis by the value $-H_{res}$. In the case of initially equal barriers, the $I_c(H)$ maximum for both directions of the transport current is at the same field value, $H=-H_{res}$ (Fig. 5(c)). The measured critical current is independent of the transport current direction, but it will be different for opposite directions of the applied magnetic field. Graphically, this is expressed in the fact that the set of curves $I_c(H)$ is symmetric about the $H$ axis.

Figure 5(d) demonstrates the most general situation, which is a combination of cases (b) and (c), i.e., there is a frozen magnetic field $H_{res}\neq 0$, and barriers are initially different. Maxima of the $I_c(H)$ dependence for the opposite current directions will be observed at different values of the field $H_-$ and $H_+$; however, in this case, $|H_+|\neq |H_-|$. Their algebraic half-sum determines the residual field in the cryostat, $H_{res}=-(H_- + H_+)/2$, and the difference $(H_+ - H_-)$ characterizes the barrier asymmetry. In the experiment, in such a case, four different curves $I_c(H)$ will be obtained when the field and current signs change; accordingly, four different values of the critical current for any arbitrary fixed field $H$ are obtained.

With respect to all these schemes, it should be noted that, for different values of the field $H$, the spatial localization of the critical state can vary; however, the maximum critical current of the film on all the curves $I_c(H)$ is the same for any combinations of the field and current directions, and is determined by the "weakest" point located on one certain edge of the film.

We have discussed in details these four cases, since they, together with the results of visualization of the critical and resistive state in wide films using LTLSM, allowed us to develop a simple, generally available technique for analyzing the critical current in a film. This method no longer requires the use of a complex spatially resolving LTLSM technique and makes it possible to obtain information on the state of each of the two edge barriers in the samples under investigation and the correspondent critical currents, and also to determine residual field in the cryostat basing on the IVC set in a weak magnetic field $H$. Such information should be useful for avoiding incorrect interpretation of experimental measurements of the critical current in wide films.

The edge barrier height affects not only the critical current magnitude, but also the subsequent resistive state, up to the break point in IVCs, shifting the resistive parts of the current-voltage characteristic along the current axis.

Figure 6 shows the current-voltage characteristics of a film with an artificially created asymmetric barrier (with a small notch made on one of the edges) measured at different mutual orientations of the external magnetic field and transport current.

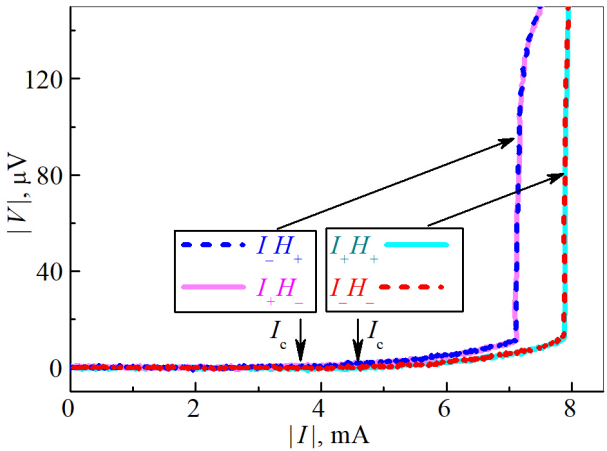

Fig. 6. I-V characteristics of a wide film strip with asymmetric edge barriers for different directions of transport current and magnetic field with induction ±0.05 mT.

Like the $I_c(H)$ dependencies in this case, I-V characteristics are grouped in pairs (compare with Fig. 5(b)). As it was shown above, the critical state arises on the "weak" edge, on which the current field and external field are added. It is evident that, due to different critical currents, the IVCs are shifted in current. Accordingly, the current value corresponding to the points of breaks (jumps) into the PSL state is numerically changed. Thus, the break-point current is determined by the vortex mechanism[24], but not by the achievement of the depairing current uniformly distributed over the cross section[22].

As the temperature is lowered, visible resistive region in the IVC corresponding to the vortex flow disappears, and transition to the PSL state becomes disruptive (Fig. 7).

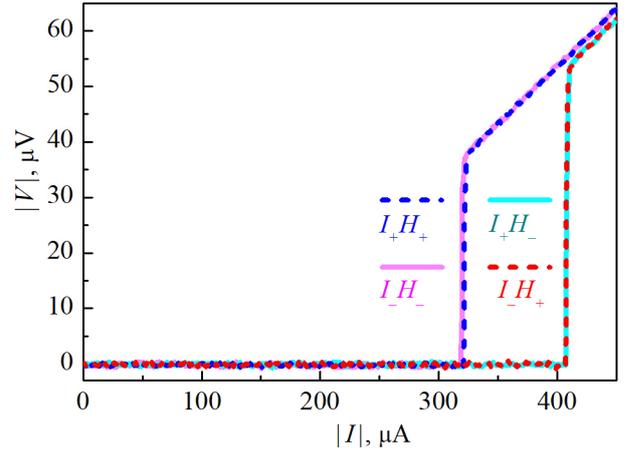

Fig. 7. Low-temperature break-off I-V characteristics of a wide film strip with asymmetric edge barriers for different directions of transport current and magnetic field of with induction ±0.06 mT.

However, discontinuous character of the I-V characteristic does not mean that the critical current ceases to be related to the penetration of the vortices into the film. Two different values of the critical current of the film at different orientations of the transport current and external magnetic field, and the pairwise coincidence of the critical current values $I_c$ for the curves $I+H+$ and $I-H-$, $I+H-$ and $I-H+$, correspondingly, show that superconductivity destruction occurs at one of the edges, and that spatial localization of the critical state may be changed by the magnetic field. This means that, even in this case, the vortex mechanism is responsible for the appearance of the critical current and resistivity, with the difference being that the vortex entering the film immediately starts to move with a velocity exceeding the Larkin-Ovchinnikov instability velocity[24].

## 5. Conclusion

The use of the space-resolution method of low-temperature laser scanning microscopy allowed us to visualize the critical and resistive state in wide tin film strips. Further, this method demonstrated that the critical state is localized at one of the edges of the strip, and that critical current is determined by the lowest of the edge barriers. The application of the external magnetic field by varying the height of the edge barriers may change the critical state localization from one edge to another. If the edges differ in their superconducting and geometric characteristics, the dependence of the critical current on the applied magnetic field for each edge may also differ. Further, for a number of practical applications (for example, superconducting film single-photon optical radiation detectors), it is important to know exactly how the critical current is determined by the edge. As superconductors such as NbN, MoSi, MoRe have very short coherence lengths, they are "wide" even at a micron and a tenth of a micron width, and it may be assumed that the results obtained on tin films simulate the behavior of the film strips of these superconductors used in single-photon detectors. In addition, it follows from the above results that, when measuring critical currents in small fields, it is necessary to pay attention to the direction of the transport current and the external perpendicular magnetic field, and to be able to change them in the experiment. On the basis of the LTLSM response maps obtained and analysis of the I-V characteristics of tin film samples in a magnetic field, a simple method for characterizing asymmetry of the edge barrier and studying each barrier separately is proposed. In addition, this method makes it possible to determine residual ("frozen") magnetic fields in the cryostat. The proposed technique does not require the use of LTLSM and includes just recording of four I-V characteristics corresponding to combinations of two directions of transport current and two opposite directions of the perpendicular magnetic field. The approach proposed in this paper enables changing of the spatial localization of the band sensitivity of superconducting electromagnetic radiation detectors and should be useful for studying various mechanisms associated with a particular edge.

___________________________________

The original paper was published in Low Temp. Phys. **44** (3), 226-232 (2018) https://doi.org/10.1063/1.5024540 (Fiz. Nizk. Temp. **44**(3), 298 (2018), in Russian).

In this post-print version you find:
- improved English translation including usage of some terms,
- the text divided in more sections (with headers 2,3,4 added),
- Figs. 3 and 4 with added color and a legend,
- extended format of the reference list (with paper titles and DOI added);
- the results are unchanged.